\documentclass[twocolumn,pre,twoside,showpacs,showkeys,floatfix, superscriptaddress]{revtex4} 

\usepackage{amsmath}
\usepackage{amssymb}
\usepackage{graphicx}
\usepackage{dcolumn}
\usepackage{bm}

\def\d{\mathrm{d}}
\def\vec#1{\mathbf{#1}}

\begin{document}

\title{Non-monotonic fluctuation spectra of membranes pinned \\ or tethered discretely to a substrate}

\author{Rolf-J\"urgen Merath}
\affiliation{
    II.\ Institut f\"ur Theoretische Physik, 
    Universit\"at Stuttgart, 
    Pfaffenwaldring 57, 
    70550 Stuttgart, 
    Germany  
}
\affiliation{
   Max-Planck-Institut f\"ur Metallforschung, 
   Heisenbergstra\ss e 3, 
   70569 Stuttgart, 
   Germany
}

\author{Udo Seifert}
\affiliation{
    II.\ Institut f\"ur Theoretische Physik, 
    Universit\"at Stuttgart, 
    Pfaffenwaldring 57, 
    70550 Stuttgart, 
    Germany  
}

\date{\today}

\begin{abstract}
The thermal fluctuation spectrum of a fluid membrane coupled harmonically to a solid support by an array of tethers is calculated. 
For strong tethers, this spectrum exhibits non-monotonic, anisotropic behavior 
with a relative maximum at a wavelength about twice the tether distance. 
The root mean square displacement is evaluated to estimate typical membrane displacements. 
Possible applications cover pillar-supported or polymer-tethered membranes. 
\end{abstract}

\pacs{82.70.-y, 87.16.Dg}
\keywords{supported membrane, fluctuation spectrum}

\maketitle

{\sl Introduction.} --- 
Thermal shape fluctuations of fluid membranes in the vicinity of a substrate depend both on the elasticity of the membrane 
and the specific type of interaction with the substrate~\cite{US_AdvPhys_1997}. 
For laterally homogeneous substrates the combination of steric, van der Waals and electrostatic interactions 
determines the strength of these fluctuations. 
Such spectra, as measured experimentally using video microscopy \cite{Zilker_1987, Raedler_1995}, 
decrease with increasing wave-vector 
since shorter wave-length fluctuations cost more bending energy. 
A qualitatively different type of interaction arises from tethering or pinning a membrane at discrete points to a substrate. 
By adjusting the length of the tethers the mean distance between substrate and membrane can be controlled. 
As tethering molecules mostly thio\-lipids are used which consist of a lipid tail, a hydrophilic spacer 
(e.\,g.\ peptides \cite{Naumann_Knoll_Langmuir_2003, Bunjes_Knoll_Langmuir_1997} 
or polymers \cite{Tanaka_2003_Wiley, Tanaka_2004}) and a sulfur-based linker to the substrate.   
Likewise, end-functionalized membrane proteins can serve as tethers~\cite{Giess_Knoll_BiophysJ_2004}. 

Micro- or nano-patterned substrates involving equi\-distant silicon pillars or gold dots 
have so far mainly been used to study the interaction with cell membranes \cite{Spatz_2004} 
or an actin cortex \cite{Spatz_2003} 
but similar experiments with vesicles should become possible as well. 
If the membrane binds specifically and firmly to these structures, it is effectively pinned at these points. 
For future biotechnological applications of such systems an understanding of how the thermal shape fluctuations are affected 
by tethering or pinning is of paramount interest. 

In this paper, we determine the spectrum of shape fluctuations of such membranes. 
Surprisingly, we find a large range of parameters for which the fluctuations exhibit a non-monotonic behaviour 
with a maximum at a wavelength of the order of the tethering or pinning distance. 
Previous theoretical work on membrane conformations on structured
substrates focussed on groove-like or rough substrates \cite{Andelman_Langmuir_1999, Andelman_PRE_2001} 
or vesicles adhering strongly to chemically patterned substrates~\cite{Lipowsky_JPhysCondMat_2005}. 
In a formally related theoretical development, 
motivated by the plasma membrane of red blood cells, fluctuations of a compound system coupling a fluid membrane to the cytoskeleton 
were investigated~\cite{Gov_2003, Fournier_2004, Gov_2004, Lin_Brown}.

\begin{figure}[!b]  
\hfil \includegraphics[scale=0.63, clip]{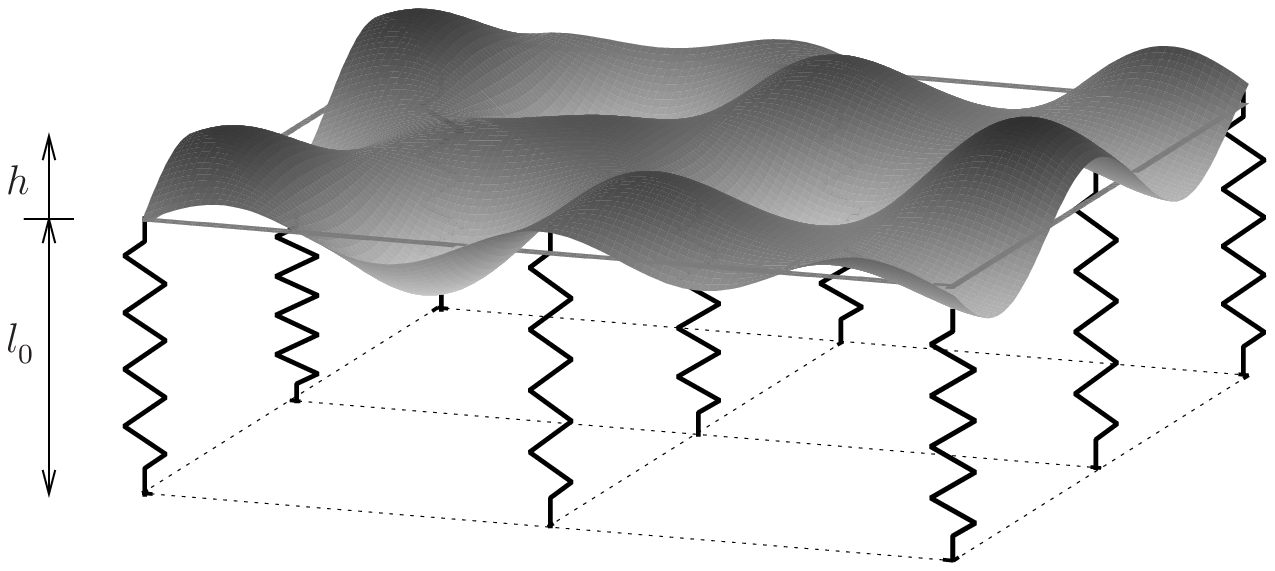}   \hfil 
\vspace{-0.2cm} 
   \caption{\label{fig:Modell-Arten}Sketch of a membrane tethered to a substrate at \mbox{discrete} sites. 
                       A membrane conformation is characterized by the height profile $h(\vec{r})$, 
                       which denotes a displacement relative to the flat configuration at a height 
                       given by the rest length $l^{~}_0$ of the tethers. 
           }       
\vspace{-0.12cm}
\end{figure}

\vspace{0.2cm}
{\sl The model.} --- 
The membrane is linked at discrete points $\vec{r}_{\alpha}$ to a substrate by $N$ springs 
as sketched in Fig.~\ref{fig:Modell-Arten}. 
The membrane surface is parameterized by a height profile $h(\vec{r}) \equiv h(x, y)$  
over a rectangular substrate of extensions $L_x$ and $L_y$. 
The height $h( \vec{r})$ refers to the displacement with respect to the rest length $l^{~}_0$ of the tethers. 
The total energy functional 
\begin{equation}  \label{eqn:E_Helfrich+Federn}
    E  \; \equiv \;  \frac{\kappa}{2} \int\limits_{0}^{L_x} \!\!   \d x  \!  \int\limits_{0}^{L_y} \!\!  \d y     \,
                     \left[\nabla^2 h(\vec{r}) \right]^2    
                     \; + \;  \sum\limits_{\alpha = 1}^{N} \frac{K_{\alpha}}{2} \;  h^2(\vec{r_{\alpha}})   
\end{equation}   
comprises membrane elasticity with bending rigidity $\kappa$
and harmonic tethers with strength $K_\alpha$.
Assuming periodic boundary conditions in the lateral direc\-tion, a Fourier expansion of the height profile reads 
\begin{equation}  \label{eqn:Fourierreihe} 
   h(\vec{r})  \: \equiv \:   \sum_\vec{k} \,h^{~}_{\vec{k}}\, e^{i\, \vec{k} \cdot \vec{r}}  
\end{equation}
with $\vec{k}  \equiv  (k_x, k_y)$ and 
$k_{x,y}  = (0, \pm 1, \pm 2, \ldots) \cdot (2 \pi / L_{x,y})$\,. 
Since $h(\vec{r})$ is real, the complex Fourier coefficients obey $h^{~}_{-\vec{k}} = h_{\vec{k}}^{*}$. 
The spatial average of the height profile, $h^{~}_\vec{0}$, is real. 
The energy can be written as      
\mbox{$E  \:  =  \:  \rule[-0.18cm]{0cm}{0.52cm}{}    
                 \frac{1}{2} \,\sum_{\vec{k},\vec{k}^\prime} \,  h^{*}_{\vec{k}} \, D^{~}_{\vec{k}, \vec{k}^{\prime}} \hspace{0.03cm} 
                                                                 h^{~}_{\vec{k}^{\prime}}  $}
with a non-diagonal coupling matrix $D^{~}_{\vec{k}, \vec{k^{\prime}}}$. 
Following a scheme introduced by Lin and Brown~\cite{Lin_Brown}, a transformation to \mbox{\emph{independent}} variables leads to 
\begin{equation}  \label{eqn:cMc}
    E ~ = ~ \vec{c} \cdot \vec{M} \, \vec{c}  
    ~ \equiv ~ \sum\limits_{r, r^\prime} c^{~}_r \:  M^{~}_{r, r^\prime}  \: c^{~}_{r^\prime}  \ \ .
\end{equation}
The components $c^{~}_r$ (with $r = 0, 1, 2, 3, \ldots$) of the vector
$\vec{c}$ can be grouped into three sectors,        
\mbox{$\vec{c} \equiv ( \frac{1}{2} h^{~}_{\vec{0}}, \;   \{ \mathrm{Re}\,h_{\vec{q}} \}  , \; \{ \mathrm{Im}\,h_{\vec{q}} \}  )$}, 
where $\vec{q}$ runs through all independent, non-vanishing wave-vectors. 
With the definitions 
\begin{equation}
   E^{~}_r \: \equiv \: \left\{ \begin{array}{ll}    0              &  \quad \mathrm{for~}r=0  \\ 
                        \kappa \, L_x L_y \, |\vec{q}(r)|^4   &   \quad \mathrm{for~}r>0   \end{array} \right.   
\end{equation}
and  
\begin{equation} 
   m_r^{\alpha} \: \equiv \: \left\{ \begin{array}{ll} 
        \!\!\!        ~~\:   \sqrt{2 K_{\alpha}}      
                                               &  ~ \mathrm{for~} r=0 ~\mathrm{(1^{st}~sector)}  \hspace*{-4cm}   \\ 
        \!\!\!    ~~\:    \sqrt{2 K_{\alpha}} \, \cos(\vec{q}(r) \cdot \vec{r}_{\alpha})  &  
                                       ~ \mathrm{for~the~2^{nd}~sector}      \hspace{-4cm}  \\     
        \!\!\! -\, \sqrt{2 K_{\alpha}} \, \sin(\vec{q}(r) \cdot \vec{r}_{\alpha})  &  ~ \mathrm{for~the~3^{rd}~sector} 
         \hspace{-4cm}   \end{array}     \right.   
\end{equation}
the matrix $\vec{M}$ 
can be simplified to take the form 
\vspace{-0.15cm}
\begin{equation}  \label{Def.MatrixM}
    M^{~}_{r,r^{\prime}} \, = \, E^{~}_r \cdot \delta^{~}_{r,r^{\prime}} 
                        \:+\: \sum_{\alpha = 1}^{N} \, m_r^{\alpha} \cdot m_{r^{\prime}}^{\alpha}   \ \ . 
\end{equation}
The fluctuation spectrum $ \rule[-0.15cm]{0cm}{0.14cm}{}   \langle \, |h^{~}_{\vec{k}}|^2 \, \rangle $ 
can then be extracted from the inverse matrix $\vec{M}^{-1}$, 
where \mbox{$M^{-1}_{r, r^{\prime}}   =   \frac{2}{k_B T} \, \langle \, c^{~}_r \, c^{~}_{r^\prime}  \rangle$~\cite{Merath}}.

Below we will compare the spectrum of the discretely tethered membrane with a simplified model called ``continuous springs'' 
where a membrane fluctuates in a laterally homogeneous 
harmonic potential of strength $\gamma$ with energy functional  
\begin{equation}  \label{eqn:E_with_gamma}
E^{\scriptscriptstyle (\gamma)}_{~}  \; \equiv \;  \frac{1}{2} 
    \int\limits_{0}^{L_x} \!\!   \d x  \!  \int\limits_{0}^{L_y} \!\!  \d y     \,
    \Big\{  \kappa  \left[\nabla^2 h(\vec{r}) \right]^2    
    \, + \,  \gamma \, h^2(\vec{r})
 \Big\}     \ \ .
\end{equation}   
The fluctuation spectrum of this system reads 
\begin{equation}   \label{eq:spectrum_gamma}
    { \langle \, |h^{~}_\vec{k}|^2 \, \rangle }^{\scriptscriptstyle (\gamma)}  
    \; = \; \frac{k_B T}{L_x L_y \, (\kappa \, |\vec{k}|^4 + \gamma)}  \ \ .
\end{equation}
The overall strength of the discrete and the continuous springs are comparable for 
$\gamma =  \frac{1}{L_x \, L_y}  \,\sum_{\alpha = 0}^{N} \, K_{\alpha}$, 
since these para\-meters lead to the 
same effective spring \mbox{constant} of the spring brush as a whole.

\begin{figure}[!tb]   
\hfil \hspace{-0.30cm} \includegraphics[scale=0.67, clip]{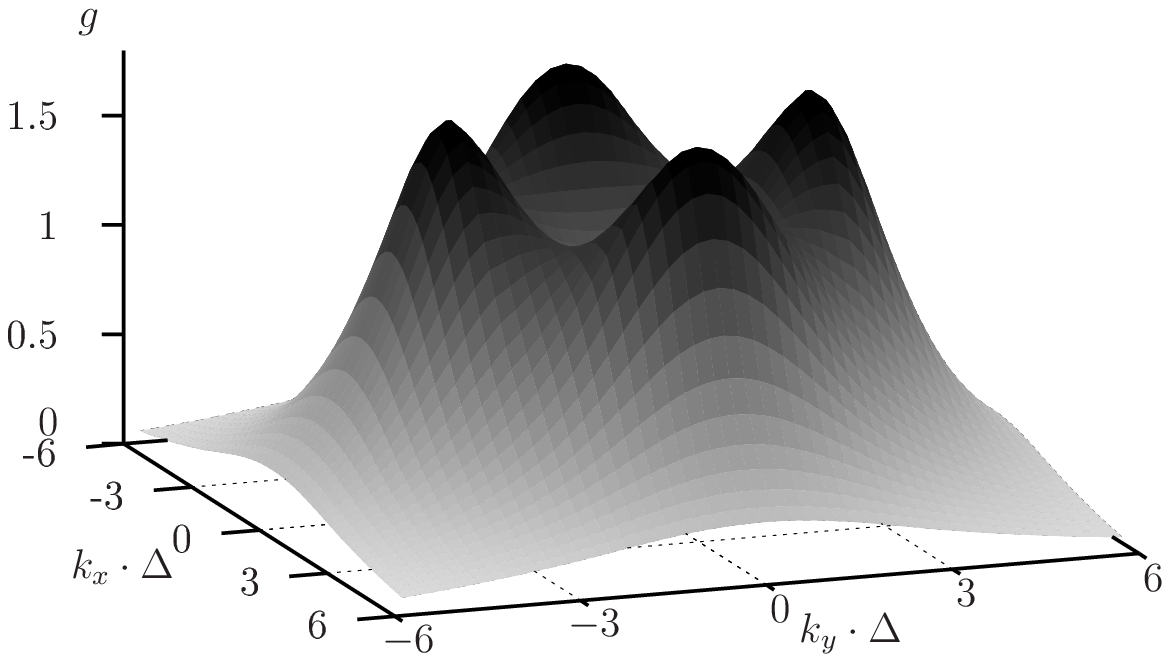} \hfil  
\vspace{-0.2cm}     
     \caption{\label{fig:quadrat._1}The scaling function $g(k_x\Delta, k_y\Delta)$ of the fluctuation spectrum 
                                    of a fluid membrane pinned by a quadratic array of infinitely strong springs. 
             }
\end{figure}

\vspace{0.2cm}   
{\sl Fluctuation spectra.} --- 
We focus on a membrane attached to a \emph{quadratic} array 
\footnote{A \emph{hexagonal} array leads to qualitatively similar results --- up to the different symmetry~\cite{Merath}.} 
of equally strong ($K_\alpha \equiv K$), equidistant tethers, with $\Delta$ as tether lattice constant, 
in the limit $L_{x, y} \to \infty$.
First, the case of \emph{pinning}, i.\,e., infinitely strong springs ($K \to \infty$), is analyzed. 
Dimensional analysis yields the spectrum in the form 
\begin{equation}  \label{eq:Def.g}
   \left.{ \langle \, |h^{~}_{\vec{k}}|^2  \, \rangle }\right|_{\mathrm{pinned}}  
       \; = \;  \left.{ \langle\, |h^{~}_{\vec{0}}|^2 \,\rangle }\right|_{\mathrm{pinned}}     \cdot  g\left(k_x\Delta, \, k_y\Delta \right) 
\end{equation}
with the amplitude at the origin ($\vec{k} = \vec{0}$) given by
\begin{equation}  \label{eq:Def.s}
   \left.{ \langle \, |h^{~}_{\vec{0}}|^2  \, \rangle }\right|_{\mathrm{pinned}} 
           \; = \; s \cdot \frac{k_B T}{\kappa \, N} \Delta^2
\end{equation}
and a numerically determined prefactor $s \simeq 3.9\cdot 10^{-3}\,$. 
The scaling function $g(k_x\Delta, k_y\Delta)$ is shown in Fig.~\ref{fig:quadrat._1}. 
The fourfold symmetry of the spectrum reflects the symmetry of the tether array. 
This spectrum is non-monotonic with four relative maxima and four saddle points. 
The wavelength at the maxima is 
$\lambda_{\mathrm{\scriptscriptstyle max}} \simeq 1.11 \cdot 2\, \Delta\,$. 
Since undulation modes with a wavelength twice the distance $\Delta$ fulfill a pinning condition 
(i.\,e.\ no displacement at the site of the tethers), 
a naive guess would yield $\lambda_{\mathrm{\scriptscriptstyle max}} =  2\, \Delta\,$. 
For a quasi-one-dimensional system (with $k_y \equiv 0$) this would indeed be true~\cite{Merath}. 
However, for the full two-dimensional case, the interplay of all fluctuation modes 
apparent in the coupling matrix $\vec{M}$ (\ref{Def.MatrixM}) 
leads to deviations from this naive expectation towards a slightly larger wavelength.

\begin{figure}[!tb]   
\hfil \hspace{-0.4cm} \includegraphics[scale=0.70, clip]{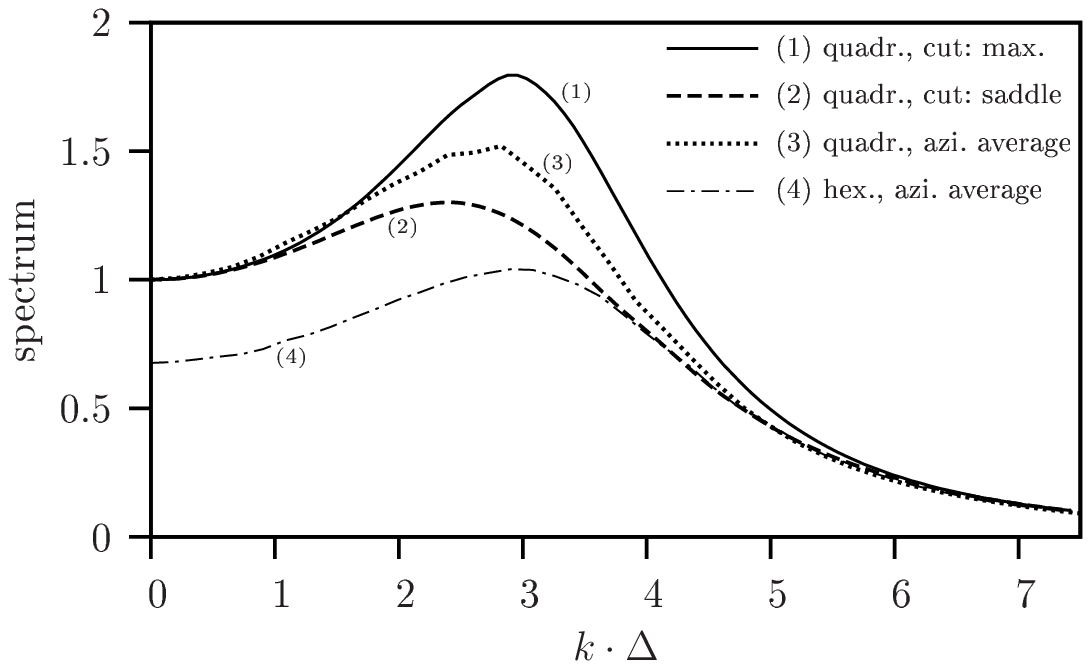} \hfil 
\vspace{-0.32cm}    
     \caption{\label{fig:cut_hard}One-dimensional plots of the universal 2-d fluctuation spectrum $g(k_x\Delta, k_y\Delta)$ 
             of a membrane pinned to a quadratic array of pinning sites (cf.\ Fig.~\ref{fig:quadrat._1}). 
             Cuts in two main directions, through a maximum (curve 1) and a saddle point (curve 2), 
             of the 2-d spectrum are shown and compared to an azimuthal average (curve 3). 
             Curve 4 shows an azimuthal average of a spectrum for a \emph{hexagonal} arrangement of pinning sites. 
             The closer packed sites induce a six-fold symmetric 2-d spectrum situated below the spectrum of the quadratic system. 
             The constant $s$ for the hexagonal pattern, 
             $s^{\mathrm{(hex.)}} \simeq  2.5\cdot10^{-3}$, 
             is different than in the quadratic case (\ref{eq:Def.s}).  
             For comparability, the hexagonal spectrum is divided by the prefactor 
             $\langle\, |h_{\vec{0}}|^2 \,\rangle^{}_{\mathrm{pinned}}$ 
             of the quadratic case (\ref{eq:Def.g}). 
             }
\end{figure}

From this spectrum, one can derive various one-dimensional spectra shown in 
Fig.~\ref{fig:cut_hard}. 
Cuts through the origin and a maximum, and through the origin and a saddle
point, respectively, show the non-monotonicity. 
The \mbox{saddle} point appears at a (by a factor of approximately $\sqrt{2}$\,) larger wavelength than the maxima, 
corresponding to the larger tether distance in this direction. 
Furthermore, the average of the 2-d spectrum with respect to the azimuth angle is displayed. 
In the averaged spectrum, the wavelength of the maximum is 
$\lambda^{\mathrm{\scriptscriptstyle azi}}_{\mathrm{\scriptscriptstyle max}} \simeq 1.2 \cdot 2\, \Delta\,$.  
It is larger than the wavelength corresponding to the relative maxi\-mum in the \mbox{2-d} spectrum 
but smaller than the wavelength corresponding to a saddle point.

\begin{figure}[!tb]
\hfil \hspace{-0.2cm} \includegraphics[scale=0.76, clip]{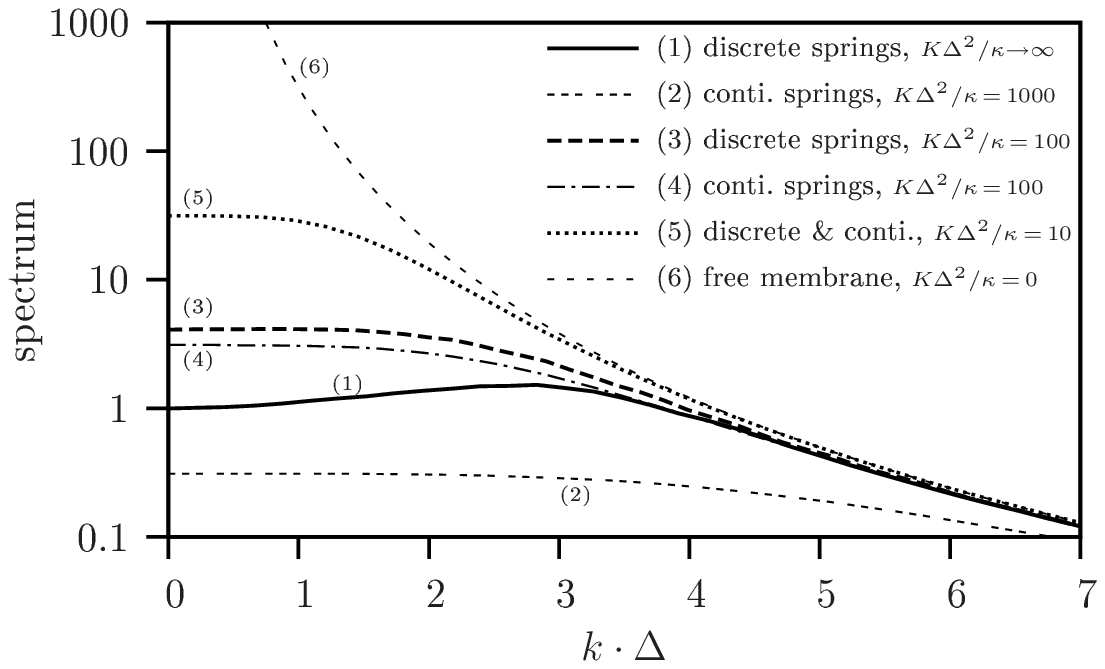} \hfil 
\vspace{-0.32cm}    
     \caption{\label{fig:Vgl.div.K}The scaling function $f(K \, \Delta^2 / \kappa, \, k_x\Delta, \, k_y\Delta)$ 
              for the fluctuation spectrum of a quadratically tethered
              membrane               for different spring stiffnesses
              (\ref{eq:Def.f}).                                                                                                                
              Azimuthal averages of the 2-d spectra are compared to results from the continuous spring model (\ref{eq:spectrum_gamma}). 
              For comparision, the spectra for the continuous model are scaled by the prefactor 
              $\langle\, |h_{\vec{0}}|^2 \,\rangle^{}_{\mathrm{pinned}}$ 
              of the discrete case.     
              Curve 1 shows the azimuthal average of the \mbox{2-d} spectrum for pinning, i.\,e.\ 
              infinitely strong springs.  
              In the comparable continuous case ($\gamma \to \infty$), the spectrum vanishes.   
              Curve 2 depicts the spectrum for strong continuous springs: 
              While the ``discrete spectrum'' is still undistinguishable from curve 1, 
              the ``continuous spectrum'' is located far underneath. 
              Curves 3 and 4 give the corresponding discrete and
              continuous spectrum for moderately strong springs, respectively. 
              For weaker springs the two cases yield nearly the same data shown in curve 5. 
              The spectrum of a free membrane without tethers is shown as curve 6. 
             }
\end{figure}

For \emph{finite spring constants $K$}, the fluctuation spectrum of a membrane tethered by a quadratic array 
can be expressed as 
\begin{equation}  \label{eq:Def.f}
   \langle \, |h^{~}_{\vec{k}}|^2  \, \rangle \; = \;   
         s \,  \frac{k_B T}{\kappa \, N} \Delta^2    \cdot f\left(\frac{K \, \Delta^2}{\kappa}, \, k_x\Delta, \, k_y\Delta\right) 
\end{equation}
with a scaling function $f$ that incorporates the spring elasticity. 
Since $f$ is continuous and $f(\infty, k_x\Delta, k_y\Delta) = g(k_x\Delta, k_y\Delta)$, 
for sufficiently large spring constants $K$ the non-monotonicity of the spectrum persists. 
In Fig.~\ref{fig:Vgl.div.K}, the wavelength-dependence of the azimuthally averaged scaling function $f$ 
is shown for different values of $K\,\Delta^2 / \kappa$  
and compared to the spectrum (\ref{eq:spectrum_gamma}) of 
a continuous confining harmonic potential. 
For weak springs, a tethered membrane behaves 
like a membrane with continuous confinement.        
For moderately strong spring constants, deviations from the continuous spectrum occur, 
with the spectrum of the discrete tethering
 being systematically larger than the continuous one. 
The positions of the relative maxima in the 2-d spectra depend on $K \Delta^2 / \kappa$. 
For infinitely strong springs the maximum of the spectrum reaches its lowest possible wavelength. 
For decreasing spring constants the wavelength corresponding to the relative maxima increases. 
Finally, below a critical value of the spring constant, 
$\left. {K \Delta^2 / \kappa  }\right|_\mathrm{crit} \, \simeq \, 100 $, 
the relative maxima disappear and the spectrum decays monotonically.

\vspace{0.2cm}  
{\sl Real-space fluctuations.} --- 
We now determine the average width of the membrane in real space. 
The height profile can be written as 
\begin{equation}
   h(\vec{r}) = 2 \, \vec{w}(\vec{r}) \cdot \vec{c} 
\end{equation}
with~\cite{Lin_Brown}
\begin{equation}
   \vec{w}(\vec{r}) \equiv ( 1, \{ \cos(\vec{q} \cdot \vec{r}) \},  \{ -\sin(\vec{q} \cdot \vec{r}) \}   ) \ \ . 
\end{equation}
Given the inverse matrix $\vec{M}^{-1}$, the mean square displacement reads 
\begin{eqnarray} \label{eq:MSD}
  \langle \, h^2(\vec{r}) \, \rangle & = & 
             4 \, \sum\limits_{r, r^\prime} w^{~}_r(\vec{r}) \: \langle \, c^{~}_r \: c^{~}_{r^\prime}  \rangle \: 
                                            w^{~}_{r^\prime}(\vec{r}) \nonumber  \\ 
             & = & 2 \; k_B T \; \sum\limits_{r, r^\prime} w^{~}_r(\vec{r}) \:  M^{-1}_{r, r^\prime}  \: w^{~}_{r^\prime}(\vec{r})   \ \ .
\end{eqnarray}
Typical membrane elongations are represented by the root mean square displacement (RMSD) $\sqrt{\langle  h^2(\vec{r}) \rangle}$. 
Only if the maximum of this RMSD is significantly smaller than the rest length of the springs, 
this model can be trusted quantitatively since it does not yet include the steric hindrance by the substrate. 
The maximum of the RMSD is a measure to estimate how far above a substrate the membrane has to be placed 
in order to avoid undesirable contact or adhesion of the bilayer 
to the substrate.

\begin{figure}[!tb]  
\vspace{0.2cm}
\hfil \hspace{-0.1cm} \includegraphics[scale=0.64, clip]{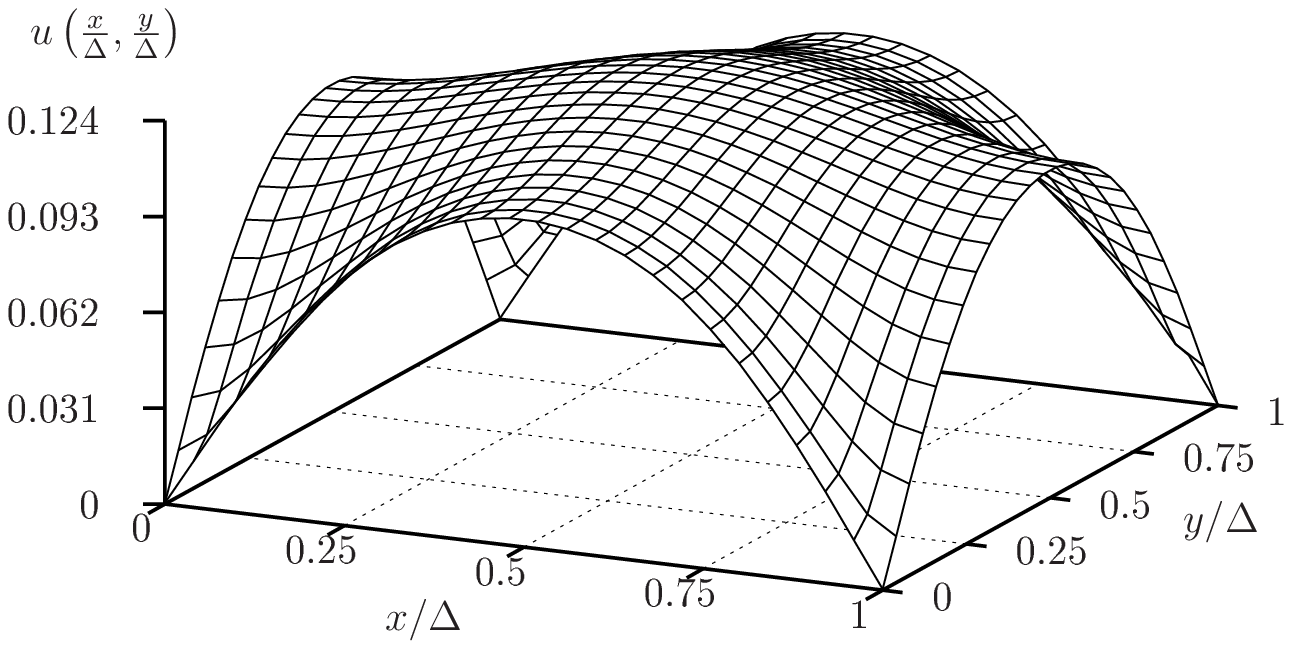} \hfil 
\vspace{-0.1cm}   
    \caption{\label{fig:Matratzen-Plot}Scaling function $u(x/\Delta, y/\Delta)$ for the RMSD, $\sqrt{\langle \, h^2(\vec{r}) \, \rangle\,} $, 
                       of a membrane pinned at a quadratic array of sites. 
            } 
\end{figure}

For \emph{pinning} ($K \to \infty$),  we find 
\begin{equation}   \label{eq:Def.u}
   \left. { \sqrt{\langle \, h^2(\vec{r}) \, \rangle\,} }\right|_{\mathrm{pinned}} \;  
           \equiv \; \sqrt{\frac{k_B T}{\kappa}\,} \Delta     \cdot  u\left(\displaystyle \frac{x}{\Delta}, \frac{y}{\Delta} \right)
\end{equation} 
with a scaling function $u(x / \Delta, y / \Delta)$     
shown in Fig.~\ref{fig:Matratzen-Plot}. 
The maximum of the RMSD in the pinned case occurs in the center of the array and is given by 
\begin{equation}  \label{eq:max_pinned}
   \left. \mathrm{max}\{ \sqrt{\langle  h^2(\vec{r}) \rangle \,} \, \}\right|_{\mathrm{pinned}} \; 
   \equiv \; p \cdot \sqrt{\frac{k_B T}{\kappa}\,} \Delta
\end{equation}
with a constant $p \simeq 0.124\,$.

For \emph{finite} spring constants $K$, the maximum of the RMSD takes the form 
\begin{equation}   \label{eq:Def.w}
   \mathrm{max}\{ \sqrt{\langle  h^2(\vec{r}) \rangle \, } \, \} \; \equiv \; p \: \sqrt{\frac{k_B T}{\kappa}\,} \Delta 
                             \cdot  w\left( \frac{K \Delta^2}{\kappa} \right) 
\end{equation} 
with a scaling function $w$.     
For weak springs, $w(K \Delta^2 / \kappa)$ approaches 
$\frac{1}{p \, \sqrt{8\,}}  \left( \frac{\kappa}{K \, \Delta^2}  \right)^{1/4}$, 
which follows from the known result 
\begin{equation}
     { \langle  h^2(\vec{r}) \rangle }^{\scriptscriptstyle (\gamma)} 
               =   \,  \frac{k_B T}{8 \sqrt{\kappa \, \gamma\,}}
\end{equation}
for a membrane bound in a harmonic potential~\cite{Safran_1994}. 
For strong springs, $w$ \mbox{approaches} 1.  
The cross-over between these two scaling limits occurs around $K\Delta^2 / \kappa = (p \, \sqrt{8\,})^{-4} \simeq 66.1$, 
where the two asymptotes intersect.

\vspace{0.4cm}
{\sl Summarizing perspective.} --- 
Discrete tethering of a membrane to a substrate has a profound implication on the fluctuation spectrum. 
For strong enough tethers, this spectrum becomes non-monotonic with a maximum determined by the spacing of the tethers. 
An interpretation of such a spectrum in terms of a continuous model would 
require a term $- |\sigma| (\nabla h)^2$ in (\ref{eqn:E_with_gamma}) implying 
a negative ``surface tension''.     
For sufficiently weak tethers, however, the discrete tethering can indeed be replaced by a continuous harmonic confining potential. 
The explicit introduction of an additional regular (positive) surface tension, 
e.\,g.\ caused by the area constraint of a vesicle whose bound part is considered~\cite{US_PRL_1995}, 
into the energy (\ref{eqn:E_Helfrich+Federn}) of our model poses no problems. 

Our quantitative data on the mean displacements caused by these fluctuations should provide valuable hints for the experimentalists, 
how large the rest length $l^{~}_0$ of the pillars and tethers has to be in
order to avoid contact with the substrate. 
For a smaller $l_0$, one should 
include a direct interaction with the substrate 
which could be attractive due to van-der-Waals interaction and/or
 repulsive due to steric interactions. 
In \mbox{either} case one could include an effective potential $V(h)$ to the energy functional. 
Its mini\-mization will then lead to a laterally inhomogeneous ``ground state'' profile $l^{~}_0(\vec{r})$ 
as in studies of membrane adhesion to structured substrates~\cite{Andelman_PRE_2001}. 
In another extension of our model one can study the fluctuation of two almost parallel membranes connected by polymeric linkers. 
The fluctuation of the relative distance between the membranes, i.\,e., the peristaltic mode, 
is then governed by an effective Hamiltonian (\ref{eqn:E_Helfrich+Federn}) 
and will show the non-monotonic behavior for strong enough linkers as well.


\end{document}